\documentclass[12pt]{article}
\usepackage{graphicx}
\usepackage{endfloat}

\begin{document}
\title{ Implementation of dense coding using the generalized Grover's algorithm}
\author{\small{} Jingfu Zhang,$^{1}$  Zhiheng Lu,$^{1}$ Lu Shan,$^{2}$ and Zhiwei
Deng$^{2}$  \\
\small{} $^{1}$Department of Physics,\\
\small{}Beijing Normal University, Beijing,
100875, People's Republic of China\\
\small{} $^{2}$Testing and Analytical Center,\\
\small{}  Beijing Normal University,
 Beijing, 100875, People's Republic of China}
\date{}
\maketitle
\begin{center}
\begin{minipage}{120mm}
\hspace{0.5cm} {\small
  Dense coding has been implemented using the generalized Grover's
algorithm and its inverse operation. Exploiting the superpositions
of two Einstein-Podolsky-Rosen (EPR) states, messages that are
possible to be transmitted increase. Our scheme is demonstrated
using nuclear magnetic resonance (NMR). Experimental results show
a good agreement between theory and experiment. }

 PACS number(s):03.67
\end{minipage}
\end{center}
\vspace{0.3cm}

\section*{1.Introduction}
    Dense coding was proposed by Bennett, and Wiesner in 1992 \cite{1}. It
has been implemented in experiments \cite{2,3}. Utilizing
entanglement properties, dense coding can transmit more than one
bit of information by manipulating only one of the two particles
in an Einstein-Podolsky-Rosen (EPR) state. Using the four EPR
states, 2 bits of information can be transmitted in the following
quantum communication scheme. Initially, Alice and Bob each obtain
one particle in a starting EPR state. Bob manipulates his particle
via one of four unitary operators so as to put the two-particle
system into one of the four EPR states and then return the treated
particle to Alice. Since the four manipulations result in the four
EPR states, 2 bits of information can be sent from Bob to Alice.
By determining the EPR state, Alice can read the encoded
information. K. Shimizu et al proposed a scheme for enhancing the
information capacity using a new degree of freedom \cite{15}. In
fact, the entangled states used in dense coding are not confined
within the EPR states. The other entangled states, such as the
superpositions of two EPR states, can also be used in dense
coding. The generalized Grover's algorithm is used to prepare the
starting EPR state, and its inversion is used to determine the
entangled state in decoding measurement.

   In our previous work, we have synthesized EPR states using
the generalized Grover's algorithm and demonstrated the
experimental results using nuclear magnetic resonance (NMR)
 \cite{4}. In this paper, we will implement dense coding using the
algorithm and demonstrate the scheme using NMR.

\section*{2.Synthesizing the superpositions of EPR states}

For convenience, the four EPR states are denoted by
$|\psi_{1}>=(|\uparrow\uparrow>+|\downarrow\downarrow>)/\sqrt{2}$,
$|\psi_{2}>=(|\uparrow\uparrow>-|\downarrow\downarrow>)/\sqrt{2}$,
$|\psi_{3}>=(|\uparrow\downarrow>+|\downarrow\uparrow>)/\sqrt{2}$,
and
$|\psi_{4}>=(|\uparrow\downarrow>-|\downarrow\uparrow>)/\sqrt{2}$,
where $|\uparrow>$ and $|\downarrow>$ are the two spin states of a
spin 1/2 particle. $|\psi_{1}>$, $|\psi_{2}>$, $|\psi_{3}>$,
$|\psi_{4}>$ are a complete orthonormal set, and arbitrary state
of the two-particle system can be expanded in terms of them. For
an example, $|\uparrow\uparrow>=(|\psi_{1}>+|\psi_{2}>)/\sqrt{2}$.

\subsection *{A.Manipulation of one particle in EPR states}
    The manipulation of particle 2 is chosen as
$R_{y}^{2}(\theta)=e^{i\theta I_{y}^{2}}$, which is expressed by

\begin{equation}\label{1}
  R_{y}^{2}(\theta)=\left(\begin{array}{cc}
    \cos(\theta/2) & \sin(\theta/2) \\
    -\sin(\theta/2) & \cos(\theta/2) \
  \end{array}\right),
\end{equation}
where $I_{y}^{2}$ denotes the matrix for $y$ component of the
angular momentum of the spin, with 2 specifying the affected
particle and by setting $\hbar=1$. For particle 2,
$R_{y}^{2}(\theta)$ transforms its states $|\uparrow>$ and
$|\downarrow>$ to the superpositions described by

\begin{equation}\label{2}
R_{y}^{2}(\theta)|\uparrow>=|\uparrow>\cos(\theta/2)
  -|\downarrow>\sin(\theta/2),
\end{equation}

\begin{equation}\label{3}
R_{y}^{2}(\theta)|\downarrow>=|\uparrow>\sin(\theta/2)
  +|\downarrow>\cos(\theta/2).
\end{equation}
Using these two equations, $R_{y}^{2}(\theta)$ transforms the four
EPR states into the superpositions of EPR states, which are
expressed by
\begin{equation}\label{4}
   R_{y}^{2}(\theta)|\psi_{1}>=|\psi_{1}>\cos(\theta/2)-|\psi_{4}>\sin(\theta/2),
\end{equation}

\begin{equation}\label{5}
   R_{y}^{2}(\theta)|\psi_{2}>=|\psi_{2}>\cos(\theta/2)-|\psi_{3}>\sin(\theta/2),
\end{equation}

\begin{equation}\label{6}
   R_{y}^{2}(\theta)|\psi_{3}>=|\psi_{3}>\cos(\theta/2)+|\psi_{2}>\sin(\theta/2),
\end{equation}

\begin{equation}\label{7}
   R_{y}^{2}(\theta)|\psi_{4}>=|\psi_{4}>\cos(\theta/2)+|\psi_{1}>\sin(\theta/2).
\end{equation}

$R_{y}^{2}(\theta)$ causes rotations of the EPR states which are
shown in Fig.1, where the black vectors denote the new states
after rotations. In Fig.1(a), for example, $R_{y}^{2}(\theta)$
rotates $|\psi_{1}>$, denoted by the horizontal vector, by
$-\theta/2$ radians in the two-dimensional vector space spanned by
$|\psi_{1}>$ and $|\psi_{4}>$. Obviously, after the rotation, each
EPR state is transformed to a superposition of two EPR states. One
should note that the superposition is still the maximally
entangled state.

   Similarly, if the manipulation is chosen as
$R_{x}^{2}(\theta)=e^{i\theta I_{x}^{2}}$, one obtains that
$R_{x}^{2}(\theta)|\psi_{1}>=|\psi_{1}>\cos(\theta/2)+i|\psi_{3}>\sin(\theta/2)$,
$R_{x}^{2}(\theta)|\psi_{2}>=|\psi_{2}>\cos(\theta/2)+i|\psi_{4}>\sin(\theta/2)$,
$R_{x}^{2}(\theta)|\psi_{3}>=|\psi_{3}>\cos(\theta/2)+i|\psi_{1}>\sin(\theta/2)$,
and
$R_{x}^{2}(\theta)|\psi_{4}>=|\psi_{4}>\cos(\theta/2)+i|\psi_{2}>\sin(\theta/2)$.
$R_{x}^{2}(\theta)$ also causes rotations of the EPR states. For
example, it rotates $|\psi_{1}>$ by $\theta/2$ radians in the
space spanned by $|\psi_{1}>$ and $i|\psi_{3}>$.

\subsection *{B.The generalized Grover's algorithm}

For a two-qubit system, the generalized Grover's algorithm can be
described as follows \cite{5}. The unitary operator $U$ is chosen
as $U=R_{y}^{1}(\varphi_{1})R_{y}^{2}(\varphi_{2})$, which is
represented as

\begin{equation}\label{8}
  U=\left(\begin{array}{cccc}
    c_{1}c_{2} & c_{1}s_{2} & s_{1}c_{2} &s_{1}s_{2} \\
    -c_{1}s_{2} &  c_{1}c_{2} & -s_{1}s_{2} & s_{1}c_{2} \\
    -s_{1}c_{2} & -s_{1}s_{2} & c_{1}c_{2} & c_{1}s_{2} \\
    s_{1}s_{2} & -s_{1}c_{2} & -c_{1}s_{2} & c_{1}c_{2} \
  \end{array}\right),
 \end{equation}
where $c_{k}(k=1,2)$ and $s_{k}$ are defined as $c_{k}\equiv
cos(\varphi_{k}/2)$, and $s_{k}\equiv sin(\varphi_{k}/2)$. The
basis states are arrayed as $|\uparrow\uparrow>$,
$|\uparrow\downarrow>$, $|\downarrow\uparrow>$,
$|\downarrow\downarrow>$. The conditional sign flip operators for
$|\uparrow\uparrow>$ and $|\downarrow\downarrow>$, and for
$|\uparrow\downarrow>$ and $|\downarrow\uparrow>$ can be chosen as
the same form represented as
\begin{equation}\label{9}
  I_{t}=\left(\begin{array}{cccc}
    1 & 0 & 0 & 0 \\
    0 & -1 & 0 & 0 \\
    0 & 0 & -1 & 0 \\
    0 & 0 & 0 & 1 \
  \end{array}\right).
 \end{equation}
If the predefined basis state is chosen as
$|s>=|\uparrow\uparrow>$, the condition phase shift is represented
as

\begin{equation}\label{10}
  I_{s}=\left(\begin{array}{cccc}
    -1 & 0 & 0 & 0 \\
    0 & 1 & 0 & 0 \\
    0 & 0 & 1 & 0 \\
    0 & 0 & 0 & 1 \
  \end{array}\right).
 \end{equation}
A composite operator $G$ is defined as
$G\equiv-UI_{s}U^{-1}I_{t}U$. When $U$ is chosen as
$U_{1}=R_{y}^{1}(\pi/4)R_{y}^{2}(3\pi/4)$,
$U_{2}=R_{y}^{1}(\pi/4)R_{y}^{2}(-3\pi/4)$,
$U_{3}=R_{y}^{1}(\pi/4)R_{y}^{2}(\pi/4)$, and
$U_{4}=R_{y}^{1}(\pi/4)R_{y}^{2}(-\pi/4)$, $G$ is represented as
$G_{j}(j=1,2,3,4)=-U_{j}I_{s}U^{-1}_{j}I_{t}U_{j}$, which are
expressed by
\begin{equation}\label{11}
  G_{1}=\frac{1}{2}\left(\begin{array}{cccc}
    -\sqrt{2}  &-1 &-1 & 0 \\
    0 & -1 & 1 & -\sqrt{2} \\
    0 & -1 & 1 & \sqrt{2} \\
    -\sqrt{2} & 1 & 1 & 0 \
  \end{array}\right),
 \end{equation}

\begin{equation}\label{12}
  G_{2}=\frac{1}{2}\left(\begin{array}{cccc}
    -\sqrt{2}  &1 &-1 & 0 \\
    0 & -1 & -1 & -\sqrt{2} \\
    0 & 1 & 1 &-\sqrt{2} \\
    \sqrt{2} & 1 & -1 & 0 \
  \end{array}\right),
 \end{equation}

\begin{equation}\label{13}
  G_{3}=\frac{1}{2}\left(\begin{array}{cccc}
    0 &1 &1 & \sqrt{2} \\
    -\sqrt{2} & 1 & -1 & 0 \\
    -\sqrt{2} & -1 & 1 & 0 \\
    0 & 1 & 1 & -\sqrt{2} \
  \end{array}\right),
 \end{equation}

\begin{equation}\label{14}
  G_{4}=\frac{1}{2}\left(\begin{array}{cccc}
    0 &-1 &1 & -\sqrt{2} \\
    \sqrt{2} & 1 & 1 & 0 \\
    -\sqrt{2} & 1 & 1 & 0 \\
    0 & 1 & -1 & -\sqrt{2} \
  \end{array}\right).
 \end{equation}
$G_{j}$ transform $|\uparrow\uparrow>$ to the four EPR states,
where the initialization step is $U_{j}|\uparrow\uparrow>$
\cite{4}. We apply $G_{j}$ to other basis states, i.e., the
initial distribution of marked and unmarked states is changed
\cite{6}, and obtain some useful results. If $G_{j}$ is applied to
$|\downarrow\downarrow>$, the initialization step is
$U_{j}|\downarrow\downarrow>$, and the results are also the four
EPR states. However, $G_{j}$ transform $|\uparrow\downarrow>$ and
$|\downarrow\uparrow>$ to the superpositions of EPR states. If
$G=G_{2}$, one obtains

\begin{equation}\label{15}
G_{2}|\uparrow\uparrow>=-|\psi_{2}>,
 \end{equation}

\begin{equation}\label{16}
G_{2}|\downarrow\downarrow>=-|\psi_{3}>,
\end{equation}

\begin{equation}\label{17}
G_{2}|\uparrow\downarrow>=(|\psi_{1}>-|\psi_{4}>)/\sqrt{2},
\end{equation}

\begin{equation}\label{18}
G_{2}|\downarrow\uparrow>=(-|\psi_{1}>-|\psi_{4}>)/\sqrt{2},
\end{equation}
using Eq. (12). Table 1 shows the results of the applications of
$G_{j}$ to various basis states. In Table 1, the top row lists the
basis states, the left column lists $G_{j}$, and the others are
results, in which the irrelevant overall phase factors can be
ignored. The superpositions of EPR states in Table 1 can be
obtained through rotating EPR states using Eqs.(4)-(7) by setting
$\theta=\pi/2$ or $-\pi/2$. For example, one can obtain Eqs.(17)
and (18) from Eq.(4), by setting $\theta=\pi/2$ and $-\pi/2$,
respectively.

\begin{center}
\begin{minipage}{100mm}
\vspace*{0.4cm}
  {\bf TABLE 1.} Results of the applications of $G_{j}(j=1,2,3,4)$ to
various basis states.
\end{minipage}
\vspace*{0.4cm}

\begin{tabular}{|c|c|c|c|c|}
\hline  &$|\uparrow\uparrow>$&$|\uparrow\downarrow>$
            &$|\downarrow\uparrow>$&$|\downarrow\downarrow>$\\
\hline
 $G_{1}$ & -$|\psi_{1}>$ &(-$|\psi_{2}>$-$|\psi_{3}>$)/$\sqrt{2}$
            &(-$|\psi_{2}>$+$|\psi_{3}>$)/$\sqrt{2}$&-$|\psi_{4}>$\\
  \hline
  $G_{2}$ & -$|\psi_{2}>$ &($|\psi_{1}>$-$|\psi_{4}>$)/$\sqrt{2}$
            &(-$|\psi_{1}>$-$|\psi_{4}>$)/$\sqrt{2}$&-$|\psi_{3}>$\\
  \hline
  $G_{3}$ & -$|\psi_{3}>$ &($|\psi_{1}>$+$|\psi_{4}>$)/$\sqrt{2}$
            &($|\psi_{1}>$-$|\psi_{4}>$)/$\sqrt{2}$&$|\psi_{2}>$\\
  \hline
  $G_{4}$ & $|\psi_{4}>$ &(-$|\psi_{2}>$+$|\psi_{3}>$)/$\sqrt{2}$
            &($|\psi_{2}>$+$|\psi_{3}>$)/$\sqrt{2}$&-$|\psi_{1}>$\\
\hline
\end{tabular}
\end{center}

  Because all the operators in $G$ are reversible, one can obtain
$G^{-1}=-U^{-1}I_{t}UI_{s}U^{-1}$, using $I_{t}^{-1}=I_{t}$, and
$I_{s}^{-1}=I_{s}$. $G^{-1}_{j}$ transform the results in Table 1
to the corresponding basis states. For example,
$G_{3}^{-1}(-|\psi_{3}>)=|\uparrow\uparrow>$,
$G_{3}^{-1}(|\psi_{1}>+|\psi_{4}>)/\sqrt{2}=|\uparrow\downarrow>$,
$G_{3}^{-1}(|\psi_{1}>-|\psi_{4}>)/\sqrt{2}=|\downarrow\uparrow>$,
and $G_{3}^{-1}|\psi_{2}>=|\downarrow\downarrow>$.

    Similarly, when $U$ is chosen as
$U_{1}=R_{x}^{1}(\pi/4)R_{x}^{2}(-3\pi/4)$,
$U_{2}=R_{x}^{1}(\pi/4)R_{x}^{2}(3\pi/4)$,
$U_{3}=R_{x}^{1}(\pi/4)R_{x}^{2}(\pi/4)$, and
$U_{4}=R_{x}^{1}(\pi/4)R_{x}^{2}(-\pi/4)$, $G_{j}$ transform
various basis states to the EPR states and the superpositions of
EPR states. The superpositions can also be obtained through
rotating the EPR states if the manipulation of particle 2 is
chosen as $R_{x}^{2}(\theta)$. For example, if $U$ is chosen as
$U_{1}=R_{x}^{1}(\pi/4)R_{x}^{2}(-3\pi/4)$, $G_{1}$ transforms
$|\uparrow\uparrow>$ to $|\psi_{1}>$, $|\downarrow\downarrow>$ to
 $|\psi_{3}>$, $|\uparrow\downarrow>$ to $(|\psi_{2}>+i|\psi_{4}>)/\sqrt{2}$,
and $|\downarrow\uparrow>$ to $(|\psi_{2}>-i|\psi_{4}>)/\sqrt{2}$,
in which the two superpositions can be obtained by
$R_{x}^{2}(\pi/2)|\psi_{2}>$ and $R_{x}^{2}(-\pi/2)|\psi_{2}>$,
respectively.

\section*{3.Scheme of dense coding using the generalized Grover's
algorithm}

   The scheme is shown in Fig. 2, where 1 and 2 denote two quantum
systems (qubits) with two states $|\uparrow>$ and $|\downarrow>$,
and $|xy>$ denotes the output basis states. $G$ denotes the
operation used to synthesize the starting EPR state using the
generalized Grover's algorithm. $V$ denotes one out of four
manipulations of qubit 2 so that the two qubit system lies in one
of four entangled states, including two EPR states and two
superpositions of EPR states. The four manipulations are
represented as $V_{1}=I$ (identity manipulation, i.e., nothing
being done),
$V_{2}=R_{y}^{2}(-\pi/2)\sigma_{x}=-R_{y}^{2}(\pi/2)\sigma_{z}$,
$V_{3}=R_{y}^{2}(\pi/2)\sigma_{x}=R_{y}^{2}(-\pi/2)\sigma_{z}$,
and $V_{4}=i\sigma_{y}$, where $\sigma_{x}$, $\sigma_{y}$ and
$\sigma_{z}$ are $x$, $y$ and $z$ components of Pauli operator,
and $\sigma_{x}=-iR_{x}(\pi)$, $\sigma_{y}=-iR_{y}(\pi)$, and
$\sigma_{z}=-iR_{z}(\pi)=-i e^{i\pi I_{z}}$. $G^{-1}$ denotes the
inversion of $G$, and it implements the decoding measurement of
the whole system. $G^{-1}$ transforms the entangled state to the
corresponding output state, by which Alice reads the encoded
information. In the scheme, two bits of information are
transmitted.

   Now, we discuss an example. Let $|\psi_{2}>$ be the starting
EPR state by setting $G=G_{2}$. $V_{1}$, $V_{2}$, $V_{3}$, and
$V_{4}$ transform $|\psi_{2}>$ to $|\psi_{2}>$,
$(|\psi_{1}>-|\psi_{4}>)/\sqrt{2}$,
 $(|\psi_{1}>+|\psi_{4}>)/\sqrt{2}$, and $|\psi_{3}>$,
respectively. After the application of $G_{2}^{-1}$, the
corresponding output states $|\uparrow\uparrow>$,
$|\uparrow\downarrow>$, $|\downarrow\uparrow>$, and
$|\downarrow\downarrow>$ are obtained, using the results in Table
1. Table 2 shows the correspondence between the starting EPR
states and the output states. In Table 2, the top row lists the
starting EPR states, the left column lists the four manipulations
in dense coding, and the others are the output states.
\begin{center}
\begin{minipage}{100mm}
\vspace*{0.4cm}
 {\bf TABLE 2.} Correspondence between the starting EPR
states and the output basis states.
\end{minipage}
\vspace*{0.4cm}
\begin{tabular}{|c|c|c|c|c|}
\hline       &$|\psi_{1}>$     &$|\psi_{2}>$   &$|\psi_{3}>$ &$|\psi_{4}>$\\
\hline
 $ V_{1}$  &$|\uparrow\uparrow>$ &$|\uparrow\uparrow>$
            &$|\uparrow\uparrow>$&$|\uparrow\uparrow>$\\
  \hline
$V_{2}$   &$|\downarrow\uparrow>$ &$|\uparrow\downarrow>$
           &$|\uparrow\downarrow>$&$|\downarrow\uparrow>$\\
  \hline
$V_{3}$ & $|\uparrow\downarrow>$ &$|\downarrow\uparrow>$
            &$|\downarrow\uparrow>$&$|\uparrow\downarrow>$\\
 \hline
$V_{4}$ & $|\downarrow\downarrow>$ &$|\downarrow\downarrow>$
            &$|\downarrow\downarrow>$&$|\downarrow\downarrow>$\\
\hline
\end{tabular}
\end{center}

    When the unitary operator $U$ in $G$ and $G^{-1}$
is chosen as $U=R_{x}^{1}(\pi/4)R_{x}^{2}(-3\pi/4)$, and $V$ is
chosen as $V_{1}^{\prime}=I$,
$V_{2}^{\prime}=R_{x}^{2}(\pi/2)\sigma_{z}$,
$V_{3}^{\prime}=R_{x}^{2}(-\pi/2)\sigma_{z}$,
$V_{4}^{\prime}=\sigma_{x}$, the scheme shown in Fig. 2 also
carries out the dense coding. When the starting EPR state is
$|\psi_{1}>$, for example, the output states are
$|\uparrow\uparrow>$,$|\uparrow\downarrow>$,
$|\downarrow\uparrow>$,$|\downarrow\downarrow>$, corresponding to
the four manipulations, respectively. Compared with the previous
scheme \cite{2,3}, there are four new manipulations that are
possible to be transmitted in the dense coding using the
superpositions of EPR states. They are $V_{2}$,
$V_{3}$,$V_{2}^{\prime}$, and $V_{3}^{\prime}$.

\section*{4. Demonstrating the scheme using NMR}
   Our experiments use a sample of carbon-13 labelled chloroform
dissolved in d6-acetone. Data are taken at room temperature with a
Bruker DRX 500 MHz spectrometer. The resonance frequencies
$\nu_{1}=125.76$ MHz for $^{13}C$, and $\nu_{2}=500.13$ MHz for
$^{1}H$. The coupling constant $J$ is measured to be 215 Hz. If
the magnetic field is along $z$ axis, the Hamiltonian of this
system is represented as \cite{7}

\begin{equation}\label{19}
  H=-2\pi\nu_{1}I_{z}^{1}-2\pi\nu_{2}I_{z}^{2}+2\pi J I_{z}^{1}
  I_{z}^{2}.
\end{equation}
In the rotating frame of spin $k$, the evolution caused by a
radio-frequency (rf) pulse on resonance along $x$ or $-y$ axis is
represented as $R_{x}^{k}(\varphi)$ or $R_{y}^{k}(-\varphi)$. The
pulse used above is denoted by $[\varphi]_{x}^{k}$ or
$[-\varphi]_{y}^{k}$. The coupled-spin evolution is denoted as

\begin{equation}\label{20}
  [\tau]=e^{-i2\pi J\tau I_{z}^{1} I_{z}^{2}}.
\end{equation}

  The pseudo-pure state is prepared by using spatial averaging
\cite{8}. The system in the equilibrium state is described by its
deviation density matrix \cite{9}
\begin{equation}\label{21}
  \rho_{eq}=\gamma_{1}I_{z}^{1}+\gamma_{2}I_{z}^{2},
\end{equation}
where $\gamma_{k}$ denotes the gyromagnetic ratio of spin $k$. The
following rf and gradient pulse sequence
$[\alpha]_{x}^{2}-[grad]_{z}-[\pi/4]_{x}^{1}-1/4J-[\pi]_{x}^{1,2}-1/4J-[-\pi]_{x}^{1,2}
-[-\pi/4]_{y}^{1}-[grad]_{z}$ transforms $\rho_{eq}$ to $\rho_{s}$
\cite{10}, which is represented as

\begin{equation}\label{22}
  \rho_{s}=I_{z}^{1}/2+I_{z}^{2}/2+I_{z}^{1}I_{z}^{2}.
\end{equation}
$\rho_{s}$ is equivalent to $|s>$ \cite{11}. The pulses are
applied from left to right.
$\alpha=\arccos(\gamma_{1}/2\gamma_{2})$,
 $[grad]_{z}$ denotes gradient pulse along $z$ axis, and the
symbol $1/4J$ means that the system evolutes under the Hamiltonian
$H$ for $1/4J$ time when pulses are switched off.
$[\pi]_{x}^{1,2}$ denotes a nonselective pulse (hard pulse). The
evolution caused by the pulse sequence
$1/4J-[\pi]_{x}^{1,2}-1/4J-[-\pi]_{x}^{1,2}$ is equivalent to the
coupled-spin evolution $[1/2J]$ described in Eq.(20)  \cite{12}.
$[\pi]_{x}^{1,2}$ pulses are applied in pairs each of which take
opposite phases in order to reduce the error accumulation causes
by imperfect calibration of $\pi$-pulses \cite{3}.

   we synthesize NMR analogs of EPR states (pseudo-EPR states)
using the generalized Grover's algorithm. Using Eq. (8), we
realize $U_{1}$, $U_{2}$, $U_{3}$ and $U_{4}$ by
$[\pi/4]_{y}^{1}-[3\pi/4]_{y}^{2}$,
 $[\pi/4]_{y}^{1}-[-3\pi/4]_{y}^{2}$,
 $[\pi/4]_{y}^{1,2}$, and
$[\pi/4]_{y}^{1}-[-\pi/4]_{y}^{2}$, respectively. Obviously,
$U_{1}^{-1}$, $U_{2}^{-1}$, $U_{3}^{-1}$, and $U_{4}^{-1}$ are
realized by $[-\pi/4]_{y}^{1}-[-3\pi/4]_{y}^{2}$,
$[-\pi/4]_{y}^{1}-[3\pi/4]_{y}^{2}$, $[-\pi/4]_{y}^{1,2}$, and
$[-\pi/4]_{y}^{1}-[\pi/4]_{y}^{2}$. Because
$I_{t}=I_{t}^{-1}=[1/J]$, $I_{t}$ and $I_{t}^{-1}$ are realized by
$1/2J-[\pi]_{x}^{1,2}-1/2J-[-\pi]_{x}^{1,2}$. According to Ref.
\cite{13}, $I_{s}$
 and $I_{s}^{-1}$ are realized by
$1/4J-[\pi]_{x}^{1,2}-1/4J-[-\pi]_{x}^{1,2}-[-\pi/2]_{y}^{1,2}-[-\pi/2]_{x}^{1,2}-
[\pi/2]_{y}^{1,2}$. $V_{2}$, $V_{3}$ and $V_{4}$ are realized by
$[\pi]_{x}^{2}-[-\pi/2]_{y}^{2}$, $[\pi]_{x}^{2}-[\pi/2]_{y}^{2}$,
and $[\pi]_{y}^{2}$, respectively.

  When the system lies in state $|\uparrow\uparrow>$, the carbon spectrum shown
in Fig.3(a) and proton spectrum shown in Fig.3(e) are recorded
through readout pulses $[\pi/2]_{y}^{1}$ and $[\pi/2]_{y}^{2}$,
respectively. Through calibrating the phases of the signals in the
two spectra, the two peaks are adjusted into absorption shapes.
These two signals are used as reference signals of which phases
are recorded to calibrate the phases of signals in other carbon
spectra and proton spectra, respectively, so that the phases of
the signals in the spectra are meaningful \cite{14}. Pulses
$[\pi]_{x}^{2}$, $[\pi]_{x}^{1}$ and $[\pi]_{x}^{1,2}$, transform
$|\uparrow\uparrow>$ to $|\uparrow\downarrow>$,
$|\downarrow\uparrow>$ and $|\downarrow\downarrow>$, respectively.
Figs.3 (b-d) show the carbon spectra obtained by
$[\pi/2]_{y}^{1}$, and Figs.3 (f-h) show the proton spectra
obtained by $[\pi/2]_{y}^{2}$, when the system lies
$|\uparrow\downarrow>$, $|\downarrow\uparrow>$ and
$|\downarrow\downarrow>$, respectively.

   The dense coding starts with $|\psi_{2}>$ by choosing $U=U_{2}$.
The four manipulations are chosen as $V_{1}$, $V_{2}$, $V_{3}$,
and $V_{4}$. Correspondingly, the carbon spectra are shown in
Figs.4(a-d), obtained by $[\pi/2]_{y}^{1}$, and the proton spectra
are shown in Figs.4(e-h), obtained by $[\pi/2]_{y}^{2}$. By
comparing Figs.4 (a-d) with Figs.3 (a-d), and Figs.4 (e-h) with
Figs.3 (e-h), respectively, we corroborate that the outputs are
$|\uparrow\uparrow>$, $|\uparrow\downarrow>$,
$|\downarrow\uparrow>$, and $|\downarrow\downarrow>$.

    When $U=U_{1}$, $U_{3}$ or $U_{4}$, the dense coding starts
with $|\psi_{1}>$, $|\psi_{3}>$ or $|\psi_{4}>$. The experimental
results are also consistent with the theoretical predictions. When
$U=U_{1}$, for example, manipulations $V_{2}$ and $V_{3}$
correspond to the carbon spectra shown in Figs.5(a-b), and the
proton spectra shown in Figs.5(c-d). The readout pulses are still
$[\pi/2]_{y}^{1}$ for the carbon spectra and $[\pi/2]_{y}^{2}$ for
the proton spectra. By comparing Figs. 5(a-b) with Figs. 3(c) and
(b), and Figs. 5(c-d) with Figs. 3(g) and (f), we corroborates
that the outputs are $|\downarrow\uparrow>$ and
$|\uparrow\downarrow>$.

\section*{4.Discussion}
 Two sets of messages can be transmitted by
choosing two kinds of $U$ in decoding measurement. One kind takes
form $R_{y}^{1}(\varphi_{1})R_{y}^{2}(\varphi_{2})$, such as
$R_{y}^{1}(\pi/4)R_{y}^{2}(3\pi/4)$. The other kind takes form
$R_{x}^{1}(\varphi_{1})R_{x}^{2}(\varphi_{2})$, such as
$R_{x}^{1}(\pi/4)R_{x}^{2}(-3\pi/4)$. The two kinds of $U$
correspond the set of manipulations containing $I$,
$R_{y}^{2}(\pi/2)\sigma_{z}$, $R_{y}^{2}(-\pi/2)\sigma_{z}$, and
$i\sigma_{y}$, and the set containing $I$,
$R_{x}^{2}(\pi/2)\sigma_{z}$, $R_{x}^{2}(-\pi/2)\sigma_{z}$, and
$\sigma_{x}$, respectively. We propose an easy scheme to transmit
the two sets of messages by introducing an ancilla two-state
system, such as a classical bit or qubit. Alice and Bob can
distinguish the two sets through the two states of the ancilla
system. The starting EPR state is chosen as $|\psi_{1}>$, for
example, by choosing $U=R_{y}^{1}(\pi/4)R_{y}^{2}(3\pi/4)$ or
$R_{x}^{1}(\pi/4)R_{x}^{2}(-3\pi/4)$, noting that the two forms
are equivalent in synthesizing $|\psi_{1}>$. Bob manipulates
particle 2 via one of unitary operators in the two sets, and
manipulates the ancilla system to denote which set the operator is
in. Then he sends particle 2 and the ancilla system to Alice.
Alice first determines the set in which the message lies by
measuring the ancilla system. According to the measurement result,
she chooses the proper kind of $U$ to read the encoding
information through the decoding measurement as shown in Fig. 2.
Because the ancilla system is independent of the two original
particles in the entangled state, the scheme is easy to carry out.

\section*{5.Conlusion}
  We combine quantum communication with the quantum search
algorithm, and propose a scheme of dense coding using the
generalized Grover's algorithm and its inversion. The experimental
results verify our scheme. The superpositions of EPR states are
used, and the messages possible to be transmitted increase.

  This work was partly supported by the National Nature Science
Foundation of China. We are also grateful to Professor Shouyong
Pei of Beijing Normal University for his helpful discussions on
the principle of quantum algorithm.

\newpage

\newpage
{\begin{center}\large{Figure Captions}\end{center}
\begin{enumerate}
\item Rotations of EPR states caused by manipulating particle 2,
one EPR particle. The black vectors denote the states after
manipulations, and they are superpositions of EPR states. The
superpositions can be obtained by rotating the horizontal EPR
states by $-\theta/2$ radians as shown in (a-b) and by $\theta/2$
radians as shown in (c-d).

\item Scheme of dense coding using the generalized Grover's
algorithm. $G$ denotes the operation carrying out the algorithm to
synthesize the starting EPR state. $V$ denotes one of the four
manipulations of particle 2. $G^{-1}$ , the inversion of $G$,
implements the decoding measurement of the whole system.

\item Carbon spectra as shown in the left column obtained through
selective readout pulse for $^{13}C$ denoted by $[\pi/2]_{y}^{1}$
and proton spectra as shown in the right column obtained through
selective readout pulse for $^{1}H$ denoted by $[\pi/2]_{y}^{2}$
when the two-spin system lies in various pseudo-pure states. The
amplitude has arbitrary units. Figures 2(a-d) and Figs. 2(e-h)
correspond to states $|\uparrow\uparrow>$, $|\uparrow\downarrow>$,
$|\downarrow\uparrow>$, and $|\downarrow\downarrow>$,
respectively. The signals in Figs. 2(a) and (e) are used as
reference signals to adjust other spectra.

\item Carbon spectra as shown in the left column obtained through
$[\pi/2]_{y}^{1}$ and proton spectra as shown in the right column
obtained through $[\pi/2]_{y}^{2}$ after the completion of dense
coding when the starting pseudo-EPR state is $|\psi_{2}>$. Figs.
4(a-d) and Figs. 4(e-h) correspond to manipulations $V_{1}$,
$V_{2}$, $V_{3}$, and $V_{4}$, respectively. Through comparing
Figs. 4(a-d) with Figs. 3(a-d) and Figs. 4(e-h) with Figs. 3(e-h),
one can corroborate that the corresponding four output states are
$|\uparrow\uparrow>$, $|\uparrow\downarrow>$,
$|\downarrow\uparrow>$, and $|\downarrow\downarrow>$.

\item Carbon spectra as shown in the left column obtained through
$[\pi/2]_{y}^{1}$ and proton spectra as shown in the right column
obtained through $[\pi/2]_{y}^{2}$ after the completion of dense
coding when the starting pseudo-EPR state is $|\psi_{1}>$. Figs.
5(a-b) and Figs. 5(c-d) correspond to $V_{2}$ and $V_{3}$,
respectively. Through comparing Figs. 5(a-b) with Figs. 3(c) and
(b), and Figs. 5(c-d) with Figs. 3(g) and (f), one can corroborate
that the two output states are $|\downarrow\uparrow>$ and
$|\uparrow\downarrow>$, respectively.

\end{enumerate}

\begin{figure}{1}
\includegraphics[width=5in]{fig1.eps}
\caption{}
\end{figure}
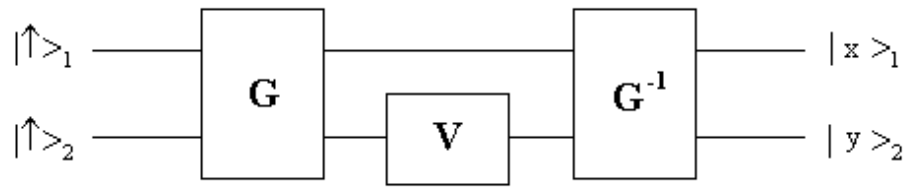
\begin{figure}{2}
\includegraphics[width=5in]{fig2.eps}
\caption{}
\end{figure}
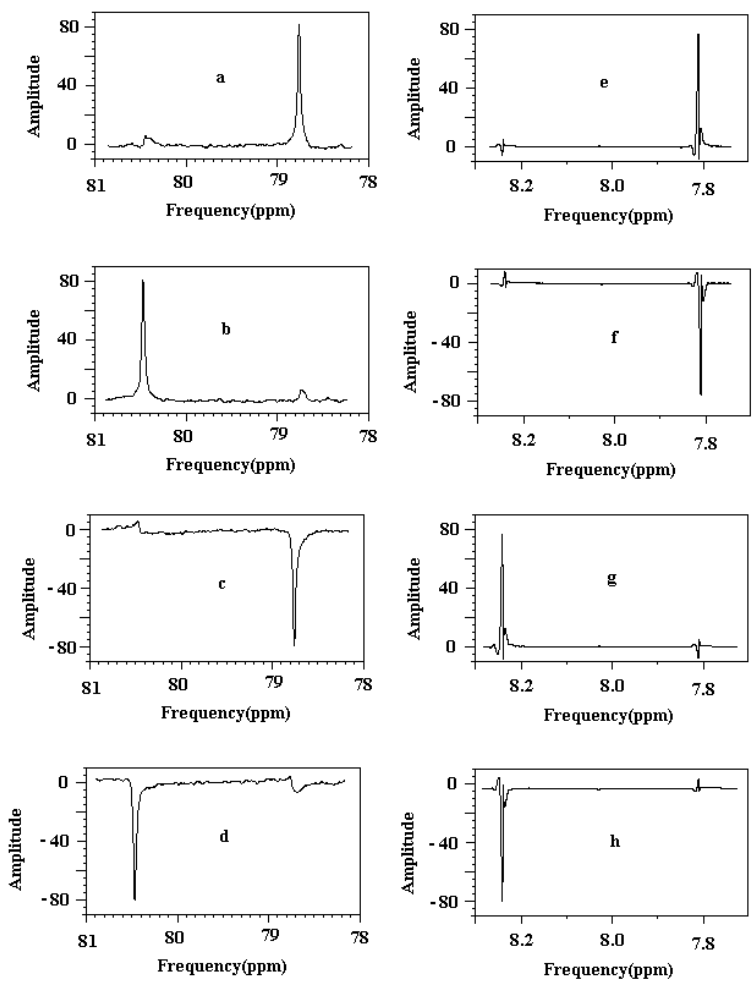
\begin{figure}{3}
\includegraphics[width=5in]{fig3.eps}
\caption{}
\end{figure}
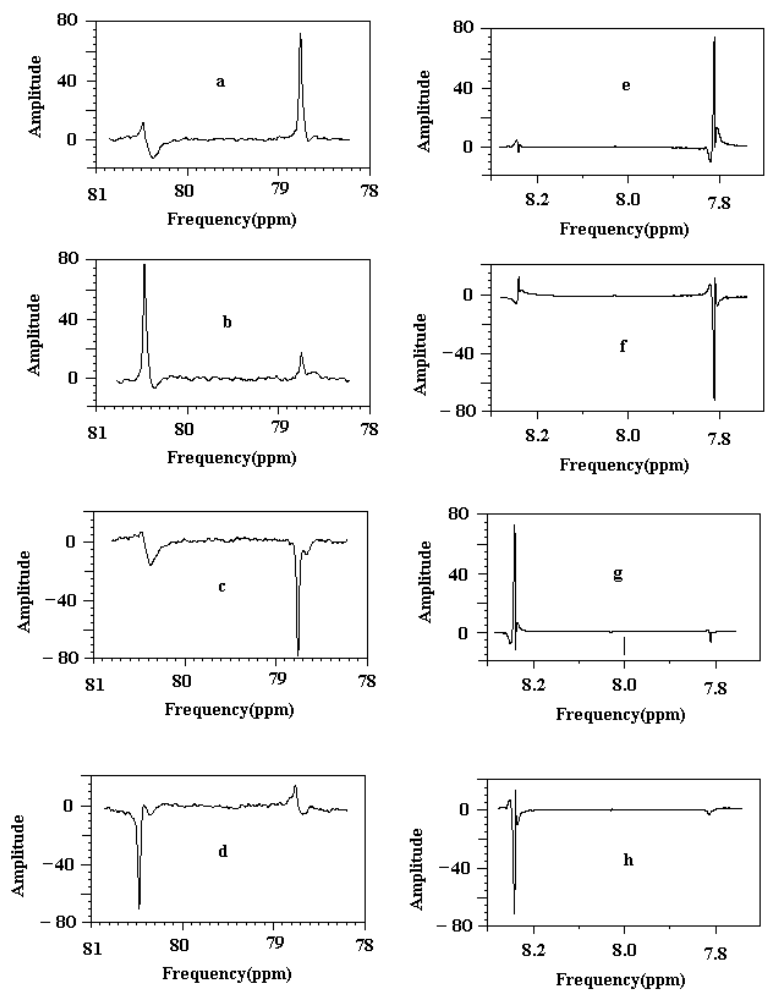
\begin{figure}{4}
\includegraphics[width=5in]{fig4.eps}
\caption{}
\end{figure}
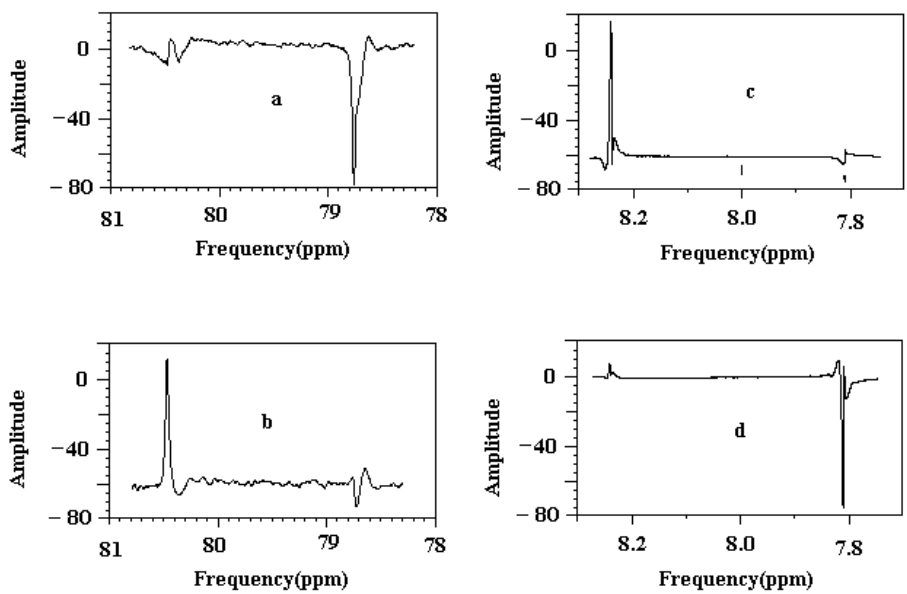
\begin{figure}{5}
\includegraphics[width=5in]{fig5.eps}
\caption{}
\end{figure}
\end{document}